%% file: main.tex
\documentclass[%
nofootinbib,
reprint,twocolumn,amsmath,amssymb,superscriptaddress,aps,prd
]{revtex4-2}

\usepackage{graphicx}
\usepackage{dcolumn}
\usepackage{bm}
\usepackage[colorlinks=true]{hyperref}
\usepackage[mathlines]{lineno}
\usepackage{aas_macros}
\usepackage{color}
\usepackage{bbm}
\usepackage{multirow}
   
\input{defs}

\begin{document}


\title{Joint reconstructions of growth and expansion histories from stage-IV surveys with minimal assumptions. II. Modified gravity and massive neutrinos.}

\author{Rodrigo~Calderón\orcid{0000-0002-8215-7292}}
 \email{calderon@kasi.re.kr}
\affiliation{Korea Astronomy and Space Science Institute,  Daejeon 34055, South Korea}
\affiliation{Laboratoire Charles Coulomb, Universit\'e de Montpellier \& CNRS, 34095 Montpellier, France}

\author{Benjamin~L'Huillier\orcid{0000-0003-2934-6243}}
\email{benjamin@sejong.ac.kr}
 \affiliation{Department of Physics and Astronomy, Sejong University, Seoul 05006, South Korea
}%

\author{David~Polarski\orcid{0000-0002-7049-8276}}
\email{david.polarski@umontpellier.fr}
\affiliation{Laboratoire Charles Coulomb, Universit\'e de Montpellier \& CNRS, 34095 Montpellier, France}

\author{Arman Shafieloo\orcid{0000-0001-6815-0337}}
\email{shafieloo@kasi.re.kr}
\affiliation{Korea Astronomy and Space Science Institute, 
 Daejeon 34055, South Korea}
\affiliation{University of Science and Technology, Daejeon 34113, South Korea}

\author{Alexei~A.~Starobinsky\orcid{0000-0002-8946-9088}}
\email{alstar@landau.ac.ru}
\affiliation{L. D. Landau Institute for Theoretical Physics RAS, Chernogolovka, Moscow region 142432, Russia}
\affiliation{Bogolyubov Laboratory of Theoretical Physics, Joint Institute for Nuclear Research, Dubna 141980, Russia}

\date{\today} 

\begin{abstract} 
Based on a formalism introduced in our previous work, we reconstruct the phenomenological function $G_{\rm eff}(z)$ describing deviations from general relativity (GR) in a model-independent manner.
In this alternative approach, we model $\mu\equiv \Geff/G$ as a Gaussian process 
and use forecasted growth-rate measurements from a stage-IV survey to reconstruct its shape for two different toy-models. 
We follow a two-step procedure: (i) we first reconstruct the background expansion history from supernovae (SNe) and baryon acoustic oscillation (BAO) measurements; (ii) we then use it to obtain the growth history $\fseight$, that we fit to redshift-space distortions (RSD) measurements to reconstruct $\Geff$. 
We find that  
upcoming surveys such as the Dark Energy Spectroscopic Instrument (DESI) might be capable of detecting deviations from GR, provided the dark energy behavior is accurately determined. We might even be able to constrain the transition redshift from $G\to\Geff$ for some particular models. 
We further assess the impact of massive neutrinos on the reconstructions of $\Geff$ (or $\mu$) assuming the expansion history is given, and only the neutrino mass is free to vary. Given the tight constraints on the neutrino mass, and for the profiles we considered in this work, we recover numerically that the effect of such massive neutrinos do not alter our conclusions.
Finally, we stress that incorrectly assuming a \lcdm\ expansion history leads to a degraded reconstruction of $\mu$, and/or a non-negligible bias in the ($\Omo$,$\seighto$)-plane. 
\end{abstract}

\maketitle

\section{Introduction}

Addressing the late-time accelerated phase of expansion of the Universe remains a major challenge for fundamental physics \cite{Weinberg:1988cp,Sahni:1999gb}. Though most observations to date are in agreement with the standard (concordance) model of cosmology (\lcdm), alternative explanations for dark energy (DE)---other than a cosmological constant $\Lambda$---are still up for debate (see e.g. \cite{perivolaropoulos2021challenges}). In particular, modifying the laws of gravity (beyond Einstein's GR) at large-scales remains a tantalizing possibility \cite{MG2012,Tsujikawa:2010zza}. 
Besides the exact nature of the dark energy (DE) component and its (effective) equation of state, additional modifications come with the properties of the relativistic degrees of freedom, notably the neutrino sector.
Interestingly, despite the wide class of modified-gravity (MG) scenarios explored in the last decades, observations seem to suggest that GR remains our best description of gravitational interactions, where dark energy is in the form of a cosmological constant in the Einstein field equations. 
For example, the detection of GW 170817, together with its electromagnetic counterpart GRB 170817A \cite{Abbott_2017}, implies that gravitational waves travel at the speed of light---ruling out a large subclass of Horndeski models predicting a tensor speed $c_T\neq c$ at the present epoch~\cite{Baker_2017}. 
Hence the detection of gravitational waves (GW) has added stringent constraints on modified gravity models in addition to  local constraints. Note that a viable cosmic expansion history can give additional strong constraints, for example, on $f(R)$ models~\cite{PhysRevLett.98.131302}.\footnote{Viable cosmological models of the present Universe in $f(R)$ gravity satisfying these constraints were independently constructed soon after that paper in~\cite{2007PhRvD..76f4004H,Appleby:2007vb,Starobinsky:2007hu}.}
At the phenomenological level, most modified theories of gravity predict a time (and possibly scale) dependent \emph{effective} gravitational coupling $G_{\rm eff}(z)~$\cite{2000PhRvL..85.2236B,De_Felice_2011} entering the equation for the growth of perturbations. 
Thus, detecting a deviation from Newton's constant would be a smoking gun for physics beyond \lcdm\ and even beyond GR. 

Let us present now the basic formalism of our approach, starting with the background. 
We consider here spatially flat Friedmann-Lemaître-Robertson-Walker universes with 
\begin{equation}
h^2(z)\equiv H^2/H_0^2=\Omo(1+z)^3+(1-\Omo)\fde(z)~, \label{h2}
\end{equation}
where $\fde=\rho_{\rm DE}(z)/\rho_{\rm DE}(z=0)$. While the second term in \eqref{h2} becomes generically subdominant in the past for viable cosmologies, this has to be enforced explicitly at high redshifts (where no data are available) once we use Gaussian processes in order to reconstruct $h(z)$ \cite{Calderon2022}. We stress further that the parameter $\Omo$ refers to \emph{clustered} dustlike matter only. The second term of \eqref{h2} is more general than the compact notation suggests, see the discussion given in \cite{Calderon2022}.
We turn now to the perturbations. 
We use the following conventions and notations \cite{2000PhRvL..85.2236B} (see also 
e.g. \cite{2013MNRAS.429.2249S})  in the conformal Newtonian gauge, where the perturbed FLRW metric is described by ($c=1$)
\begin{align}
\diff s^2 = -(1+2\phi) \diff t^2 + (1-2\psi) a^2 \diff \vect x^2,
\end{align}
where $\phi$ and $\psi$ are the Bardeen potentials. 
Phenomenologically, on subhorizon scales, in many modified gravity models the departure from the standard perburbations growth in GR is encoded in 
the  modified  Poisson  equation  \cite{2000PhRvL..85.2236B} (see also e.g.~\cite{2010PhRvD..81j4023P, 2013MNRAS.429.2249S, 2016PhRvD..94j4014P})
\begin{align}\label{poisson}
\nabla^2 \phi = 4\pi G_\text{eff}(a,\vect{k}) ~\rho \equiv 4\pi G \mu(a,\vect k)  ~\rho~.
\end{align}  
GR corresponds obviously to $\mu \equiv 1$. 
The relation between the Bardeen potentials is expressed as follows 
\begin{align} 
\phi \equiv \eta(a,\vect k) ~\psi~,
\end{align}
the two potentials are generically unequal in these models.
The subhorizon modes are essentially affected by $\mu$ as is explicit from Eq. \eqref{eq:growth} given below, while super horizon modes are affected by both $\mu$  and 
$\eta$ \citep{2010PhRvD..81j4023P}. 
In this work, given the datasets considered, we restrict our attention to $\mu$ (see e.g.~\cite{2018JCAP...11..027M,mu_GA} for constraints on $\eta$).
In what follows, we will use \Geff\ and $\mu$ interchangeably, since $\mu$ is just \Geff\ in units of $G$.
The growth of dustlike subhorizon matter perturbations in the quasi-static Approximation (QSA) is then governed by \cite{2000PhRvL..85.2236B}
\begin{equation}
\ddot\delta + 2H \dot\delta = 4\pi G\, \rho\, \delta\, \mu(z,\vect{k}), \label{eq:growth}
\end{equation}
where $\delta\equiv\delta\rho/\rho$ is the density contrast of dustlike matter. For modes of cosmological interest, the $k$-dependence of $\mu$ is often mild and can be neglected in a first approach \cite{PhysRevD.86.123503,PhysRevD.89.083505,Bose_2015}---see e.g.~\cite{Baker_2014,Johnson_2016,Garcia_Quintero_2020,2022JCAP...11..029D} for current and future constraints on the scaledependence of $\mu$. 
Note that this is certainly the case for the unscreened scalar-tensor model considered in \cite{2000PhRvL..85.2236B}.
We will restrict ourselves here to phenomenological models where $\mu$ or $G_\text{eff}$ is scale independent.\\

The above equation can be rewritten in terms of the growth factor $f\equiv\delta'/\delta$, to give
\begin{align}\label{eq:growth_f}
f' + \left(f+2+\frac{h'}{h}\right) f - \frac 3 2 \Omm(z) \mu(z) & = 0~, 
\end{align}
where a prime stands for derivative with respect to $N\equiv \ln{a}$. 
From an observational standpoint, redshift space distortions (RSD) provide us with growth rate measurements of the quantity  
\begin{equation}
f\sigma_8\equiv\frac{\sigma_{8,0}}{\delta_0} f\delta = \frac{\sigma_{8,0}}{\delta_0}\delta',
\;\;\;\text{ with }\;\;\;
\delta_0=\delta(z=0)~.
\end{equation}
We remind that the quantities $\Omega_i$ appearing in \eqref{h2} and \eqref{eq:growth_f} are defined in the standard way as in GR with the help of Newton's constant $G$. 

In this work, we will use the synergy between geometrical background probes (type Ia supernovae [SN] and baryon acoustic oscillations [BAO]) and growth measurements from RSD to constrain the phenomenological function $\mu(z)$ describing the departures from GR.
While current analysis pipelines rely on various assumptions (namely, $\Lambda$+GR) when extracting the cosmological information from large-scale structure, in particular the BAO and RSD measurements, we expect that our results will remain essentially unaffected when such effects are taken into account.

The paper is organized as follows. We start by describing in detail the methodology and the data used in Sec.~\ref{sec:Method_Data}. In Sec.~\ref{sec:Results}, we apply the method to simulated RSD data generated with $\mu\neq1$ in both idealistic and realistic scenarios and further discuss the implications of the results. We also comment on the effects of incorrectly assuming a \lcdm\ expansion history on the reconstructions in Sec.~\ref{sec:lcdm_bkg}. 
In Sec.~\ref{sec:mnu}, we consider separately the inclusion of massive neutrinos. 

\section{Method and Data}\label{sec:Method_Data}

\subsection{Models and Mock Data}
For the data, we generate mock 
$\fseight$ measurements for a (stage-IV) DESI-like survey following Tables 2.3--2.7 in \cite{DESI:2016fyo} (covering $14$  K $\deg^2$) and for different behaviors of \Geff\ that we aim to reconstruct. 
Namely, we consider an $f(R)$-inspired  \emph{bumplike} profile (which we refer to simply as ``bump'') and a smooth \emph{steplike} transition (``dip'' hereafter) in the recent past towards the weak gravity regime ($\Geff<G$), see e.g. \citep{2021mgca.book..507K,Gannouji:2020ylf}.\footnote{Indeed, both such profiles can occur in viable cosmological models in $f(R)$ gravity, see \cite {Motohashi:2010tb} in particular, especially in the case of oscillations around phantom divide~\cite{Motohashi:2011wy}.} 
These two profiles are treated purely phenomenologically here,\footnote{{Ref. \cite{Saridakis:2023pzo} presented a concrete MG model with similar profiles for $\mu(z)$ considered in this work (or rather their reflections along the $\mu=1$ axis), which could simultaneously ease the $H_0$ and $\sigma_8$ tensions.}} indeed viable $f(R)$ theories are actually screened and allow $G_{\rm eff,0}$ to deviate from $G$ today. 
Nonetheless, due to the $k$-dependence of $\mu$ which we do not discuss here, cosmic scales smaller than some critical scale would experience a boost in their growth in the recent past. 

In the case of the dip, we consider it mainly to assess whether such profiles can be accurately reconstructed using our model-independent approach.
Note in this context that a decreasing $\mu$ is impossible in massless scalar-tensor models \cite{Gannouji:2018ncm}.
To summarize, these hybrid profiles allow us to test our reconstruction independently of any theoretical prior. 
\begin{figure}
    \centering
    \includegraphics[width=\columnwidth]{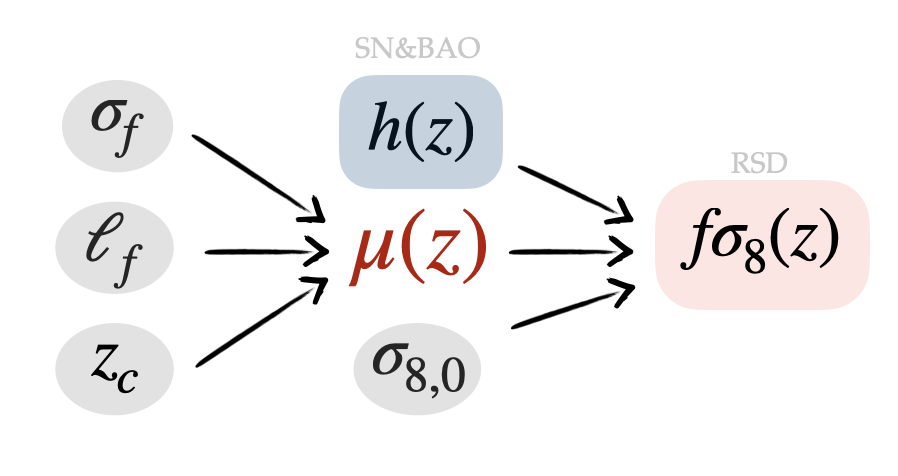}
    \caption{{Schematic representation of our analysis pipeline. We sample the (log) of the hyperparameters ($\sigma_f,\ell_f,z_c$) to obtain samples of $\mu(z)$, which we use -- together with a given expansion history $h(z)$ and a value of $\sigma_8$ -- to compute the growth history $\fseight(z)$ that we compare to DESI-like RSD measurements to ``reconstruct'' $\mu(z)$. The expansion history is reconstructed in a model-independent fashion, using the methodology introduced in \cite{Calderon2022}.}}
    \label{fig:diagram}
\end{figure}
The behaviors of the phenomenological functions $\mu^{\rm fid}(z)$ used to generate the data are depicted by the dashed lines in the upper panel of Fig. \ref{fig:ideal_joint}, while the corresponding growth $\fseight(z)$ evolutions are shown in the lower panel.
We also make use of stage-IV SN+BAO data to determine the background expansion history $h(z)$ without relying on a specific parametric model, as explained in Sec.~\ref{sec:Real}. The fiducial background used to generate the data is a Chevallier-Polarski-Linder (CPL) model \cite{2001IJMPD..10..213C,2003PhRvL..90i1301L}, extensively discussed in \cite{Calderon2022} with
\begin{multline}\label{eq:fid}
    \theta^{\mathrm {fid} }=\{\Omega_{\mathrm{m,0}} ^{\mathrm{fid}}=0.28,w_0^{\mathrm {fid}}=-0.95,w_{a}^{\mathrm {fid}}=0.3, \\ 
    h_\mathrm{fid} = 0.7,
     \sigma_{8,0}^{\mathrm {fid}}=0.81\},
\end{multline}
where $H_0 = 100\, h\, \si{km.s^{-1}.Mpc^{-1}}$.
More details on the background-only (SN+BAO) mock data can also be found in \cite{Calderon2022}. Already at this stage, let us note that modified theories of gravity can lead to a modified Chandrasekhar mass (with $m_{\rm ch}\sim \Geff^{-3/2}$ \cite{Amendola99}), relevant for SNeIa analyses, which can affect the absolute magnitude [{e.g.} $\Delta M=\frac{15}{4}\log \mu(z)$ in scalar-tensor theories\footnote{Note however that this theoretical correction can be even smaller, if the stretch correction is taken into account \cite{linderprivate}} \cite{2001PhRvD..65b3506G,Gannouji_2006})]] and hence the distance measurements obtained from such standard candles \cite{Wright_2018,Zhao:2018gwk,Ballardini21}. This effect has even been proposed as a possible explanation for the mismatch between early and late-time measurements of the Hubble constant $H_0$, see e.g.~\cite{Benevento:2020fev,Braglia:2020iik,Ballardini:2020iws,Alestas:2020zol,Pogosian:2021mcs,Perivolaropoulos:2022khd,Alestas:2022xxm}. 
However, for our purposes, we neglect these effects and assume the $h(z)$ measurements obtained from SNe are independent of $\mu$ in the current analysis. The inclusion of these effects for a specific model might be the subject of future works.

\subsection{The method}
\begin{figure}
    \centering
    \includegraphics[width=\columnwidth]{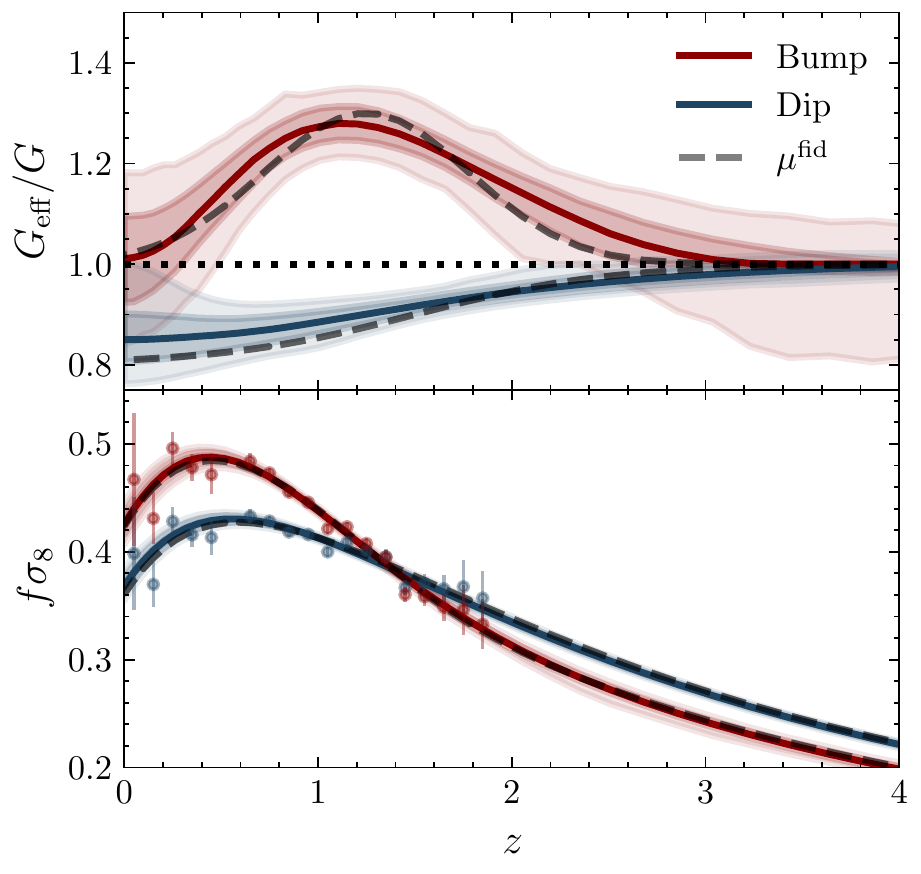}
    \caption{Reconstructions of \Geff\ in the idealistic case where the background $h(z)$ and amplitude of fluctuations $\sigma_{8,0}$ are perfectly known. Solid lines and shaded regions correspond to the median, $68\%$ and $95\%$ confidence intervals around it, respectively. Dashed lines correspond to the fiducial cosmologies generating the DESI-like (RSD) data. The redshift $z_c$ of the transition to GR, as well as the hyperparameters $\sigma_f$ and $\ell_f$ appearing in \eqref{Geff} are nonetheless  free parameters to be determined by the data.  Both of these reconstructions detect deviations from GR ($\mu=1$) at more than $2\sigma$ for $z\sim1$.}
    \label{fig:ideal_joint}
\end{figure}

To explore possible modifications of gravity at late times, we model $\Geff(z)$ as a Gaussian process\footnote{We do not delve into the details of Gaussian process modeling here, instead we refer the reader to our previous work \cite{Calderon2022} and the excellent review \cite{rasmussen2006gaussian} for more. Note that in this work, unlike common notations in the GP literature, $\mu$ refers to the phenomenological function appearing in \eqref{poisson}, and the mean of the GP is denoted by $\bar{f}$.} (GP) centered around Newton's constant $G$, such that 
\begin{equation}
     \mu(z;\sigma_f,\ell_f,z_c)=
    \begin{cases}
      \mathcal{GP}(\bar{f}(z)=1,k(\sigma_f,\ell_f)), & \text{for}\ z<z_c\\
      1, & \text{for}\ z\geq z_c
    \end{cases}
    \label{Geff}
\end{equation}
so that we recover GR at large-$z$. 
We ``pretrain'' our GP with the following theoretical priors: 
\begin{subequations}
\begin{align} \label{eq:pretrain}
    \mu(0) & = 1\pm 0.05, \\
    \mu(z_c) & =1 \pm 10^{-10},\\
    \mu'(0) & = \mu'(z_c) =0 \pm 10^{-10}.
\end{align}
\end{subequations}
These conditions allow us to smoothly recover $\Geff=G$ above a certain $z_c$ and at $z=0$, while exploring possible departures from GR at intermediate redshifts $0.1<z<10$ (see e.g.~\cite{2011JCAP...07..037N,Espejo:2018hxa,Raveri2021dbu,Pogosian:2021mcs,mu_GA,Ruiz-Zapatero:2022xbv,Ruiz-Zapatero:2022zpx,Heisenberg_gqk} for other approaches). 
Recovering $\Geff(0)=G$ is not strictly necessary (see our discussion at the beginning of this section), but from a technical point of view it can help guide our reconstructions at very low $z$ where we are volume limited and uncertainties become quite large. 
Furthermore, when dealing with real data, we do not know the true behavior of $\mu$, and whether the underlying model is screened or not, hence the two representative behaviors at $z=0$ chosen for our profiles.
It is comforting to find that the first condition does not alter the reconstruction of the second profile around $z=0$ as illustrated by the blue curves in Fig. \ref{fig:ideal_joint}.\\

We use a squared exponential kernel given by
\begin{equation}\label{eq:kernel}
    k(z,z';\sigma_f,\ell_f)=\sigma_f^2\,e^{-(z-z')^2/2\ell_f^2},
\end{equation}
where $\sigma_f$ and $\ell_f$ determine the amplitude and typical length scale of the correlations, respectively \cite{rasmussen2006gaussian}.\\

In a Bayesian spirit, we give flat {uninformative and wide } priors to the cosmological and (hyper)parameters, listed in Table \ref{tab:priors_params}. 
We sample the parameter space using Markov chain Monte Carlo (MCMC) methods, as implemented in \texttt{emcee} \cite{2010CAMCS...5...65G,Foreman_Mackey_2013}. At each step in the MCMC, we draw a sample of $\mu(N=\ln{a})\equiv\Geff/G\sim \mathcal{GP}(1,K)$, characterized by $(\sigma_f,\ell_f,z_c)$, and solve the growth equation {for a given value of \seighto\ and expansion $h(z)$, to obtain a solution $\fseight(z)$---see the diagram in Fig. \ref{fig:diagram}---prior to any comparison with the data.
Note that the parameter $z_c$ also enters prior to any computation of the likelihood and is irrespective of the data points.  
In other words, this
can be seen as \emph{forward-modeling}, rather than training the GP with the data in the usual sense. Thus, we rely on the maximization of the following likelihood function
\begin{equation}
    \ln{\mathcal{L}}=-\frac12~\chisq=-\frac12~\vect{r}^\intercal\cdot\tens{C}^{-1}\cdot\vect{r},
\end{equation}
where $\vect{r}=\fseight^{\rm GP} - \fseight^{\rm data}$ is the residual vector and $\tens{C}^{-1}$ is the inverse of the covariance matrix. The growth history $\fseight^{\rm GP}$ is obtained by solving the Eq. \eqref{eq:growth_f} for each ``pre-trained'' sample of $\mu(z)$ drawn from Eq. \eqref{Geff}.}
Those samples of $\mu(z)$ retracing a similar shape to $\mu^{\rm fid}$ will yield a better fit to growth data, and thus will be statistically favored in the long run. Averaging over a large number of realizations gives the median shape of $\mu(z)$ and $95\%\,(2 \sigma)$ confidence intervals around it. This is along the lines of what was done in \cite{Calderon2022} to reconstruct \fde, but this time we also include conditions 
on the derivatives of the GP, to smoothly recover the form in Eq. \eqref{Geff}, following the formalism described in Appendix \ref{GP_derivative}.

\begin{table}
    \centering
    \begin{tabular}{ccccc}
        \hline
		Parameter & $\sigma_{8,0}$ & $\log_{10}\sigma_f$ & $\log_{10}\ell_f$  & $\log_{10}z_c$ \\ 
		\hline
		Prior & $[0.5,1.2]$ & $[-3,0.5]$ & $[-1,0.2]$ & $[-1,1]$ \\ 
		\hline
    \end{tabular}
    \caption{Uniform priors for the parameters used in the MCMC analyses.}
    \label{tab:priors_params}
\end{table}

\section{Results and Discussions}\label{sec:Results}
\subsection{Ideal case: Background is perfectly known}
We first consider the idealistic case where the background expansion history is perfectly known. In other words, we fix $\Omo$ and $\seighto$ to their fiducial values and further assume that the dark energy evolution is known $\fde(z)=\fde^{\rm fid}$. Although this is far from being a realistic scenario, it allows us to test our method and quantify the uncertainties purely coming from the modifications of gravity, encoded in $\Geff$. 

The posterior distributions for $\mu(z)$ assuming perfect knowledge of $h(z)$ and \seighto\ are shown in Fig. \ref{fig:ideal_joint}. If the background (and the amplitude of fluctuations \seighto) are perfectly known, the RSD data alone are enough to perform an accurate (within $2\sigma$) reconstruction of the underlying theory of gravity, \emph{i.e.} $\Geff(z)$. In the next subsection, we take a more realistic approach,
where only minimal assumptions on the background are made\footnote{We only assume a flat FLRW universe, and that the Hubble rate is a sum of a matter term and an ``effective'' DE component \cite{Calderon2022}} and $h(z)$ is purely determined from the data.

\subsection{Realistic case: \tpdf{\seighto} free---\tpdf{$\Omo$} and \tpdf{$\fde(z)$} determined by SN+BAO}\label{sec:Real}

\begin{figure}
    \centering
    \includegraphics[width=\columnwidth]{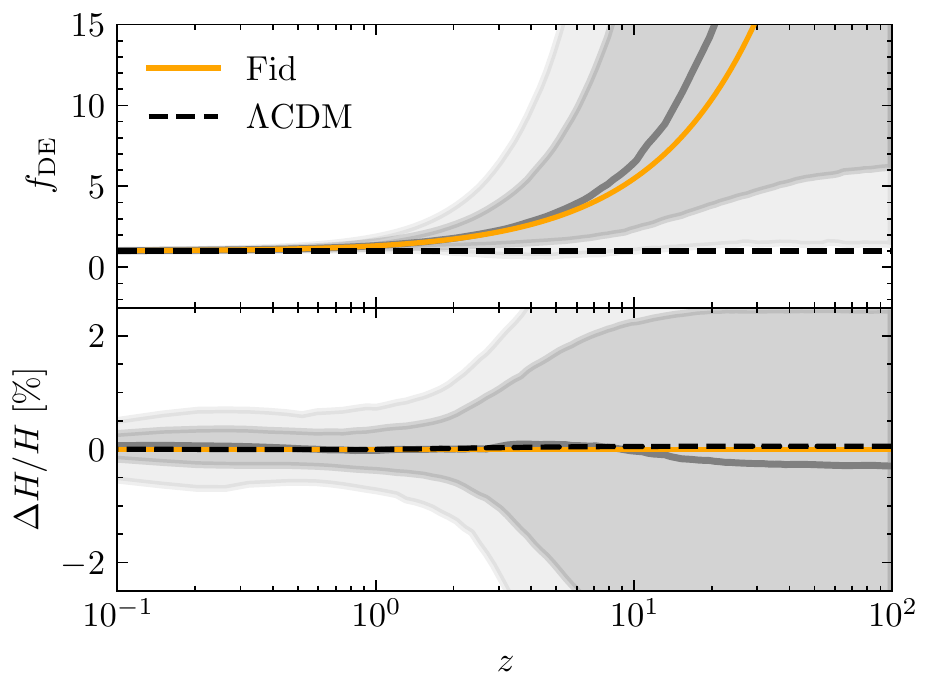}
    \caption{\emph{Top:} Reconstruction of the DE evolution $\fde(z)$. \emph{Bottom:} Relative (percentage) errors in the background reconstructions from forecasted SN+BAO measurements. The orange line corresponds to the true fiducial background in \eqref{eq:fid}, while gray lines depict the reconstructed median, 68 and 95\% confidence levels around it. Dashed-black line correspond to \lcdm's best-fit ($\fde=1$, $\Omo=0.3103$) to SN+BAO data.}
    \label{fig:bkg-reco}
\end{figure}

\begin{figure}
    \centering
    \includegraphics[width=\columnwidth]{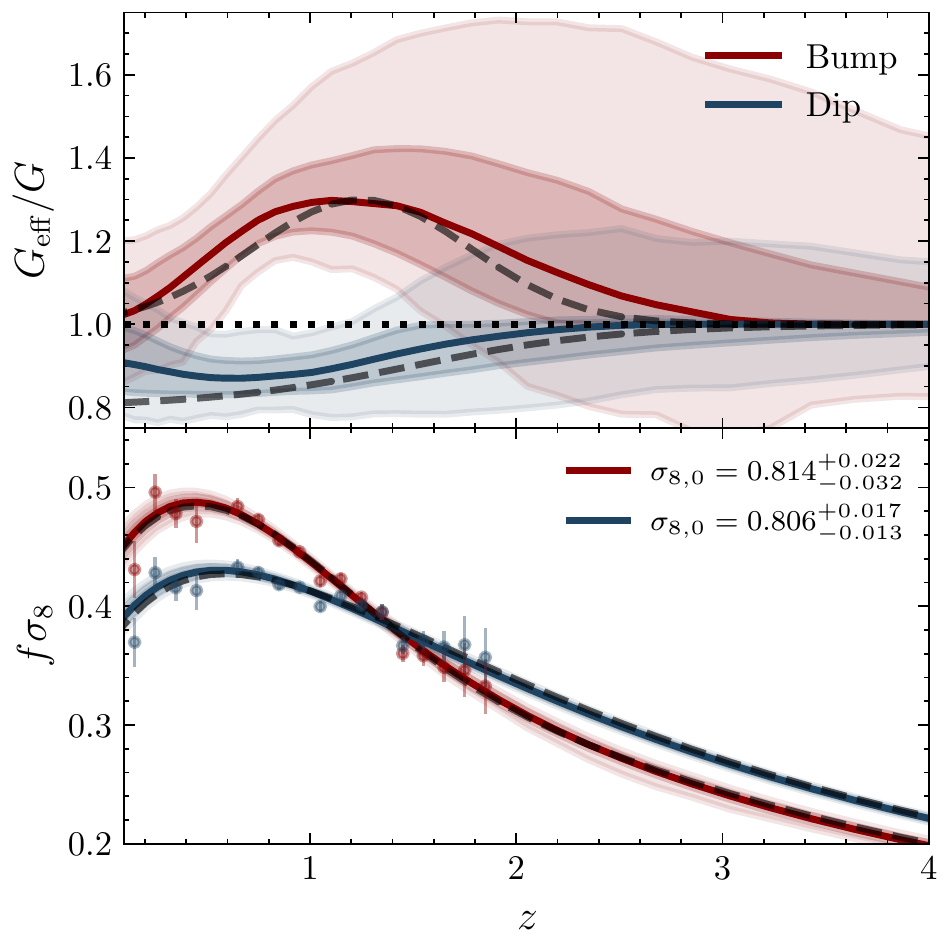}
    \caption{Realistic case where $\seighto$ is allowed to vary, and the background $h(z)$ is determined by SN+BAO (gray lines in Fig. \ref{fig:bkg-reco}). The fiducial cosmologies used to generate the $\fseight(z)$ measurements are shown by the dashed lines. Despite having larger error confidence intervals with respect to the idealistic case in Fig. \ref{fig:ideal_joint}, both of these reconstructions are still able to rule out GR at more than $2\sigma$ at $z\sim1$.}
    \label{fig:realistic}
\end{figure}

In this section, instead of assuming a parametric form for $h(z)$, we use the reconstructed expansion history as determined by SN+BAO data. In practice, this amounts to obtaining an expansion history $h(z)$ from the samples of \fde\ and calculating angular and luminosity distances which are then fitted to the data, as explained in \cite{Calderon2022}.
The degeneracies between $\seighto, \Omo$ and $\Geff$ make it very hard to say something about the underlying theory of gravity, given the quality of the data and, in particular, when all parameters are free to vary. To circumvent this issue, we assume a single expansion history, as determined solely by the data. More specifically, the expansion history $h(z)$, along with the value of \Omo--needed for solving the growth equation \eqref{eq:growth_f}--is the median of all the realizations drawn from the Markov SN+BAO chains,\footnote{The posterior distributions correspond to the blue contours shown in Fig. 6 of \citet{Calderon2022}.} obtained in \cite{Calderon2022}. Indeed, it was shown in \cite{Calderon2022} that our method is able to capture a large class of DE models, even those where the contribution from DE is not negligible at high-$z$. Our reconstruction of $h(z)$ is accurate to $\lesssim1\%$ across the entire redshift range of interest--see Fig. \ref{fig:bkg-reco}. The amplitude of the fluctuations, $\seighto$, now becomes a free parameter, and we sample the full parameter space $\theta=\{\seighto,\log_{10}\sigma_f,\log_{10}\ell_f,\log_{10}z_c\}$ in the range given by Table \ref{tab:priors_params}. In Fig. \ref{fig:realistic}, we show the reconstructions when using the median of $h(z)$ and median $\Omo$ from the SN+BAO chains. As expected, the uncertainties in the reconstructions increase with respect to those in Fig.\ref{fig:ideal_joint}, as \seighto\ is now a free parameter that is somewhat degenerate with $\Geff$, allowing for more flexibility in the samples of \Geff\ drawn at each step in MCMC.  

\begin{figure}
    \centering
    \includegraphics[width=\columnwidth]{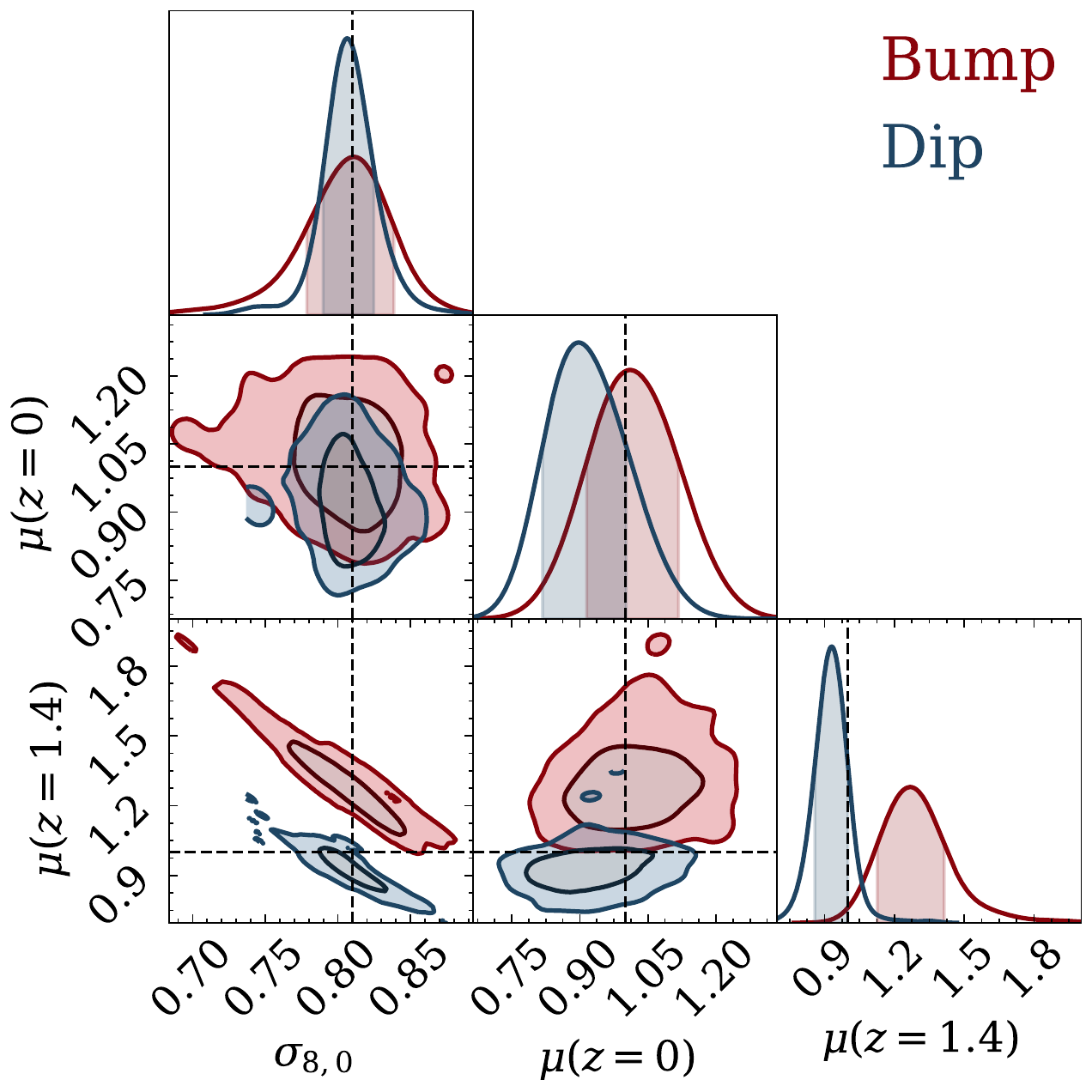}
    \caption{Marginalized posterior distributions of the relevant cosmological parameters, when using our model-independent reconstructions of $h(z)$, shown in gray in Fig. \ref{fig:bkg-reco} [where the unknown function $\fde(z)$ is reconstructed in a fully model-independent way and \Omo\ is fixed to the median of all possible values obtained from the SN+BAO chains---\emph{cf.} Sec.~\ref{sec:Real}]. }
    \label{fig:posteriors_mu_z14}
\end{figure}

The advantage of taking this approach is that we do not make any assumption on the evolution of DE, and we are able to effectively reconstruct any expansion history directly from the data, by reconstructing $\fde(z)$. Moreover, this disentangles the uncertainties coming from the growth evolution $\fseight(z)$ and those coming from the background expansion $h(z)$. This also allows us to point down a value for $\Omo$, which is of course anticorrelated with \seighto, which is in turn anticorrelated with $\Geff$. Thus, allowing for more constraining power on the quantity of interest $\mu(z)$ from RSD alone. The two-dimensional posteriors of the quantity $\mu$ at two different redshifts $z=0$ and $z=1.4$ are shown in Fig. \ref{fig:posteriors_mu_z14}. At $z=1.4$, where most of the constraining power of RSD measurements lies, the bumplike posteriors in red exclude GR ($\mu=1$, in dashed) at $>2\sigma$, while the posteriors for the diplike profile in blue are marginally consistent with GR at the $2\sigma$ level. At low redshift, because of the large uncertainties in $\fseight$, the posteriors are much broader and provide a $\sim20\%$ constraint on $\mu(z=0)$. We note that the study of peculiar velocities using SNIa from ZTF and LSST can potentially improve the measurements of the growth at very low-$z$ by a factor of 2 with respect to DESI \cite{https://doi.org/10.48550/arxiv.2001.09095}---see also \cite{Palmese_2021} for other interesting constraints using gravitational waves and galaxies' peculiar velocities.   
Interestingly, because the redshift $z_c$ in \eqref{Geff} of the transition from $G\to\Geff$ is a free parameter, our method allows us to constrain when the departures from GR start taking place
(see Fig. \ref{fig:posteriors_ST_vs_fR} and the discussions in Appendix \ref{Supplementary}). For the particular profiles considered in this work, the posterior distribution of $z_c$ is quite peaked, and we have a ``detection'' of a transition from $G
\to \Geff$ in both cases, as seen from Fig. \ref{fig:posteriors_ST_vs_fR}. The corresponding constraints on $z_c$ are given in Table \ref{tab:Constraints}

\subsection{Incorrectly assuming a \tpdf{\lcdm} background}\label{sec:lcdm_bkg}
\begin{figure}
    \centering
    \includegraphics[width=\columnwidth]{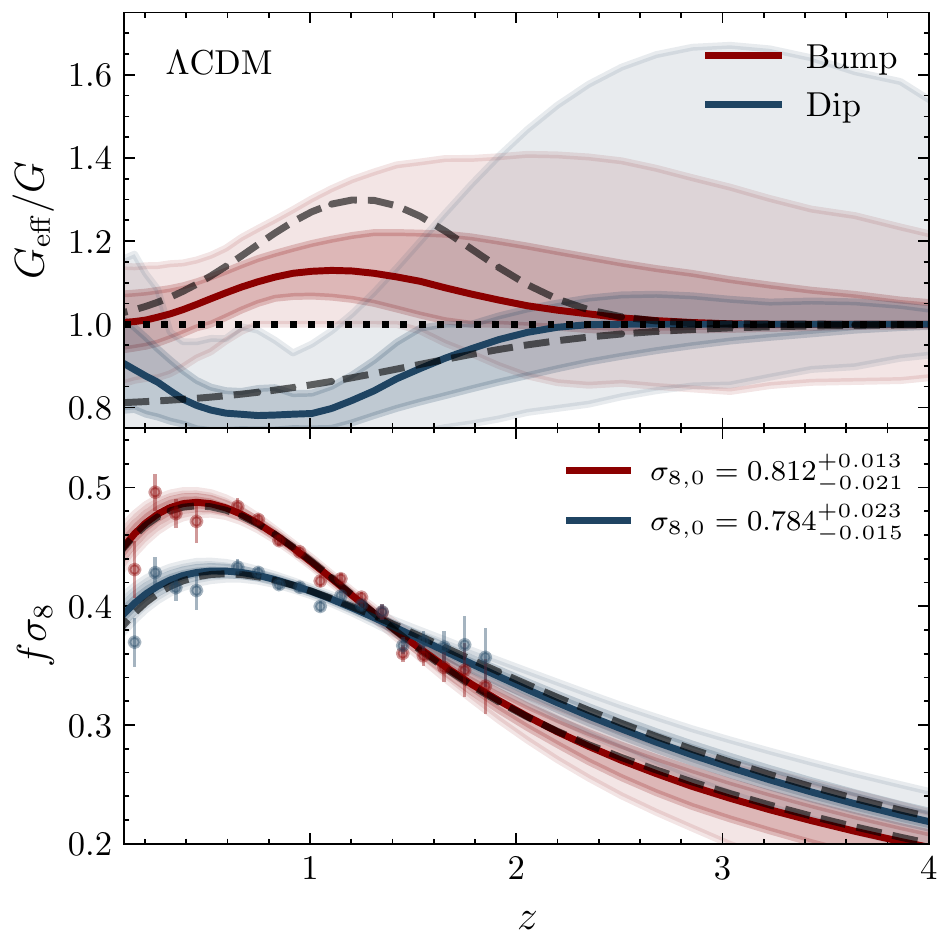}
    \caption{Reconstructions of $\mu(z)$ when assuming the best-fit \lcdm's expansion history, with $\Omo=0.31$. Incorrectly assuming a \lcdm\ background leads to biased determinations of \seighto\ and a degraded reconstruction of $\mu(z)$, despite being perfectly consistent with $\fseight(z)$, as can be seen from the lower panel. }
    \label{fig:realistic_LCDM}
\end{figure}
Cosmological observations suggest that dark energy is in the form of a cosmological constant $\Lambda$. Because of its simplicity and agreement with observations, it remains the standard model of cosmology today. Thus, most cosmological analyses are done within the \lcdm\ framework, which might lead to biased reconstructions if DE is \emph{not constant}, as for the fiducial cosmology considered here. In this section, we explore the effects of incorrectly assuming a \lcdm\ background expansion history in the reconstructions of $\mu(z)$. In other words, we fit a \lcdm\ model to the SN+BAO mock data described before and find the corresponding best-fit value for $\Omo$ (and thus $\Omega_{\Lambda,0}=1-\Omo$). We remind the reader that the mock data are generated from a time-evolving CPL dark energy model, given by Eq. \eqref{eq:fid}. We then use this expansion history to solve for the perturbations and reconstruct $\mu(z)$, as explained in the previous sections. The black dashed lines in Fig. \ref{fig:bkg-reco} show the best-fit \lcdm\ expansion history (with $\Omo^{\Lambda\rm CDM,bf}=0.3103^{+0.0025}_{-0.0024}$), compared to the fiducial one with $\Omo^{\rm fid}=0.28$ in orange (hence, representing a $\sim12\sigma$ bias in the fractional matter density). Despite having almost identical $H(z)$, the differences in the DE evolution $\fde(z)$ and biased $\Omo$ translate into a degraded reconstruction of $\Geff$, shown in Fig. \ref{fig:realistic_LCDM}---to be compared with Fig. \ref{fig:realistic}. We also find that the inferred value of $\seighto$ can be biased $\seighto\sim0.78$ vs. $\seighto^{\rm fid}=0.81$ (which corresponds to a $\sim1.2\sigma$ bias in the inferred amplitude of fluctuations) for the case of the dip (in blue)---see Table \ref{tab:Constraints}. As understood from our previous work \cite{Calderon2022}, from the background-only (SN+BAO) standpoint, the lack of DE at high-$z$ is compensated by higher values of \Omo, which translates into lower values of \seighto\ (or lower $\Geff<\Geff^{\rm fid}$) to maintain the agreement with growth-rate measurements of $f\sigma_8(z)$. This is a perfect example of what might happen if one incorrectly assumes DE is constant, the background expansion history might be consistent with the geometrical probes (SN+BAO), but a tension might appear in the amplitude of fluctuations $\seighto$ inferred from LSS observables. 
Despite the bias in the cosmological parameters \Omo\ and \seighto---and for the specific cases  of $\mu(z)$ considered here---the reconstructions are still able to capture the main trends in $\mu(z)$.\\

Finally, let us note that for the steplike transition in blue, the reason why the reconstructions deviate somehow from the fiducial $\mu^{\rm fid}(z)$ (in dashed) at very low-$z$ is because of our theoretical prior $\Geff(z=0)\simeq G$, which tends to draw our GP samples back to 1. We stress that this prior does not need to be imposed, as we do not necessarily have $\Geff(z=0)\simeq G$ in most MG theories. We have in mind here theories without screening mechanisms that do require $\Geff\simeq G$ today to satisfy local constraints, e.g. \cite{Ballardini:2021evv}. Despite this prior, because of the large uncertainties in RSD measurements at $z\sim0$, our reconstructions are still able to capture (within $2\sigma$) the true fiducial $\mu^{\rm fid}_{\rm Dip}$.

\section{Effect of massive neutrinos}\label{sec:mnu}

In this section, we consider universes containing massive neutrinos. We want to investigate how well our reconstruction of $\mu$ fares in their presence. It is well known that  free-streaming species with nonzero mass (here massive neutrinos) lead to a suppression of gravitational clustering on scales below a characteristic scale, corresponding to their free-streaming length. 
Hence, while massive neutrinos contribute to the universe's expansion in the same way as usual dustlike matter (corresponding to $\Omega_{m}$), they are absent from the driving term in the matter perturbations growth. Hence we have in front of us a situation where the parameter 
$\Omega_{{\rm m},0}$ does not represent all dustlike components at low redshifts.  
Indeed, one cannot distinguish massive neutrinos from dustlike matter purely from geometric probes at low $z$. In this case, the splitting in \eqref{h2}, while sensible theoretically, is somewhat ambiguous regarding expansion data if we have no additional information on $\Omega_{{\rm m},0}$ or $\Omega_{\nu,0}$. 
This ambiguity however gets broken once we consider the perturbations growth.
In a first step, we assume the presence of massive neutrinos and we work with equation \eqref{Hnu} below [instead of \eqref{h2}]. 
So, while we reconstruct $\mu$ as a Gaussian process, we assume the background expansion is known up to the two parameters $\Omega_{{\rm m},0}$ and $m_{\nu}$. Here however, we have only one free parameter left. 
Indeed, in this section we fix the present relative energy density $\Omo^\text{tot}$ of all components which behave like dust at low $z$, namely, 
\begin{align}
\Omo^\text{tot} \equiv \Omo+\Omega_{\nu,0} = \Omega_{\text{cdm},0}+\Omega_{\text{b},0} +\Omega_{\nu,0}~,
\end{align}
 where $\Omega_{\text{cdm},0}$, $\Omega_{\text{b},0}$, and $\Omega_{\nu,0}$ are the present relative densities of cold dark matter, baryons, and massive neutrinos respectively. Note that the couple of parameters $\left(\Omega_{{\rm m},0}, m_{\nu}\right)$ and $\left(\Omega_{{\rm m},0}^{\rm tot}, m_{\nu}\right)$ carry the same information.

We assume now that $h^2(z)$ is given by
\begin{align}\label{Hnu} 
 h^2(z) &=
 ~\Omo ~(1+z)^3 + \Omega_{\Lambda,0}\\ 
\nonumber & +\Omega_{\gamma,0}~(1+z)^4 \left( 1+0.2271\,\frac{N_\text{eff}}{3}\, \sum_i f_\nu \left(\frac{m_{\nu_i}}{T_{\nu}}\right) \right),  
\end{align}  
where $f_\nu(y) \simeq (1+(Ay)^p)^{1/p}$ is a fit provided in Ref. \cite{2011ApJS..192...18K}, with $A=\frac{180 \zeta(3)} {7\pi^4}$ and $p = 1.83$; where $\zeta$ is Riemann's $\zeta$-function.
This fitting function $f_\nu$ describes the evolution from the relativistic behavior when 
$m_\nu \ll T_\nu$ ($T_\nu\sim a^{-1}$) to the nonrelativistic regime when we have eventually 
$m_\nu \gg T_\nu$.
Like in \eqref{h2}, the first term appearing in \eqref{Hnu} corresponds to the fractional amount of matter that clusters. 
In order to test our reconstruction in the presence of massive neutrinos, it is more relevant to consider universes sharing identical $\Omega_{{\rm m},0}^{{\rm tot}}$ rather than identical $\Omega_{{\rm m},0}$, but with different $\Omega_{{\rm m},0}$, or equivalently different neutrino masses $m_{\nu}$. 
Clearly, the parameters $\Omega_{{\rm m,0}}^{\rm tot}$ and $m_\nu$, completely define the background expansion \eqref{Hnu}.

The driving term in the perturbations growth equation depends on the combination $\mu ~\Omega_m$. Hence for modified gravity and in the presence of massive neutrinos, this combination is modified at low redshifts as follows      
\begin{align}
G \Omega_{\rm m}^{{\rm tot}}\to \Geff \Omega_{\rm m}
& = G \Omega_{\rm m}^{{\rm tot}}  ~\mu \left( 1 - 
\frac{ \Omega_{\nu} }{ \Omega_{\rm m}^{{\rm tot}} } \right)  \nonumber \\ 
&
\approx 0.965 \frac {m_{\nu}}{\SI{0.5} {eV}}h_{70}^{-2}  \mu ~G  \Omega_{\rm m}^{{\rm tot}} ,
\end{align} 
where we evidently have $\Omega_{\rm m}^{{\rm tot}} = \Omega_{\rm m}$ in the absence of massive neutrinos, and $h_{70} = H_0/\SI{70}{km.s^{-1}.{Mpc}^{-1}}$. 
For the values we take here, the change comes essentially from modified gravity.

Here, we forecast the future surveys' potential to reconstruct the coupling strength $\mu(z)$ in the presence of massive neutrinos and purely from RSD measurements of $\fseight(z)$. As before, we generate mock data from a fiducial model; this time we choose a (\lcdm$\nu$) cosmology containing 2 massless and 1 massive neutrinos, with $m_{\nu}^{\rm fid}=0.5\;\rm eV$. Although this mass is larger than what is currently allowed by cosmological observations\footnote{Cosmological constraints are indirect and somewhat model dependent, unlike ground-based experiments.} \cite{Alam:2021,Planck:2018vyg}, it is still within the allowed mass range probed by terrestrial experiments, which constrain $m_{\nu}^2\equiv\Sigma_i \abs{U_{ei}}^2m_i^2= 0.26^{+0.34}_{-0.34}\;\rm eV^2$, yielding an upper bound on the electron (anti)-neutrino mass $m_\nu<0.8\;\rm eV$ at $90\%$ confidence level \cite{KATRIN:2021uub}\footnote{Note that masses of usual and sterile neutrinos $m_{\nu}\sim 1 \,\rm eV$ are well possible in viable $f(R)$ cosmological models~\cite{Motohashi:2012wc,Chudaykin:2014oia}.}. The rest of the cosmological parameters are fixed to Planck's best-fit values. Because of the growth suppression from such a massive neutrino, the normalization of the matter power spectrum $P_{\rm m}(k,z=0)$, characterized by \seighto, is now $\seighto^{\rm fid}\simeq0.73$, lower than in the previous sections (where \seighto\ was fixed to $\seighto^{\rm fid}=0.81$).\\
In what follows, we assume that this normalization ($\seighto^{\rm fid}=0.73$, as obtained for $\mu =1$) is the same for all profiles of \Geff. Although the actual normalization of the $P_{\rm m}(k,z=0)$ would indeed depend on the theory of gravity, we generate mock data for different profiles of $\mu$ from the same value of \seighto. We stress that this choice is arbitrary, as we are dealing with simulated data and we are interested in assessing whether the theory of gravity $\mu(z)$ and $\seighto$ are accurately recovered by our model-independent reconstructions, which do not know anything about the underlying theory that generates the data.

\begin{figure}
    \centering
    \includegraphics[width=\columnwidth]{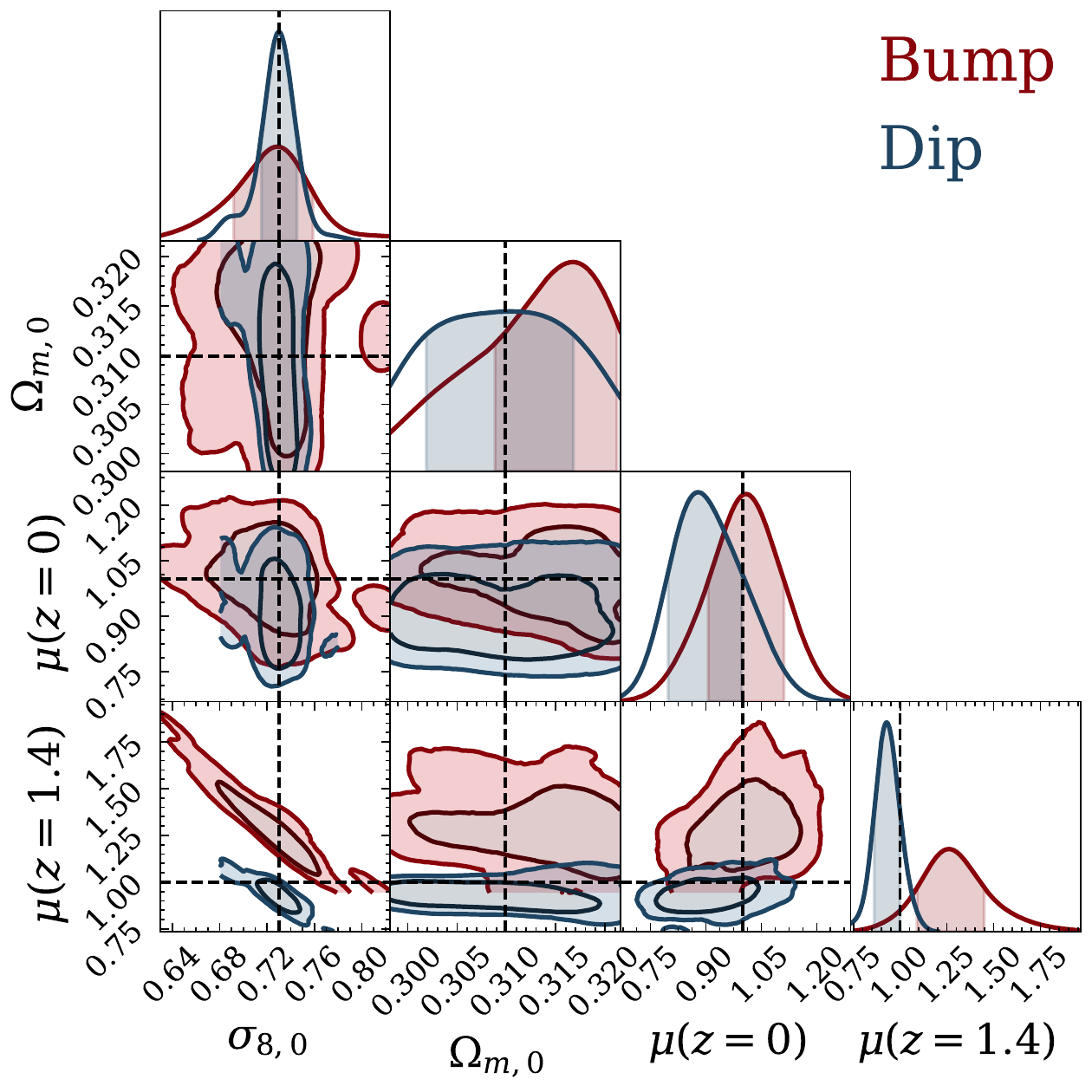}
    \caption{Marginalized posterior distributions for the parameters in the presence of massive neutrinos. This figure is the same as Fig. \ref{fig:posteriors_mu_z14}, but this time assuming the background is known $h(z)$ (up to 1 free parameter $m_\nu$) given by Eq. \eqref{Hnu}, including relativistic species and when the neutrino mass is free to vary (\emph{cf.} Sec.~\ref{sec:mnu}).}
    \label{fig:triangle_mnu_Omcl}
\end{figure}

We then sample the parameters
$\theta=\{\seighto,m_{\nu},\log_{10}\sigma_f,\log_{10}\ell_f,\log_{10}z_c\}$, with $m_{\nu}\in [0,1]$ to see the impact of a varying neutrino mass on the reconstructions of $\mu(z)$.
The posterior distributions for the relevant cosmological parameters are shown on 
Fig. \ref{fig:triangle_mnu_Omcl}. {Although we sample $m_{\nu}$, we show the posterior distributions for the derived parameter \Omo, as it corresponds to the driving term for the growth in the right-hand side of Eq.\eqref{eq:growth_f} and the actual neutrino mass is unconstrained.}
The value of \seighto\ is anticorrelated with the reconstructions of $\mu$, mainly seen in the $(\seighto,\mu(z=1.4))$-plane. Large deviations from GR, up to $\mu(z=1.4)\sim1.8$ can be achieved, provided that the amplitude of fluctuations \seighto\ is low ($\seighto\sim0.65$). A slight (negative) correlation between $\Omo$ and \seighto\ is also obtained, as expected. The enhanced suppression of growth (due to larger mass $m_{\nu}$, hence smaller $\Omo=
\Omtot-\Omega_{\nu,0}$) needs to be compensated by larger values of $\seighto$, to maintain the agreement with \fseight\ measurements. Despite these correlations, the reconstructions of $\mu(z)$ remain accurate, and does not seem to be affected by a varying neutrino mass (other than increasing the uncertainties in the reconstructions, due to an additional free parameter). The fiducial value for $\seighto$, shown as a dashed vertical line in 
Fig. \ref{fig:triangle_mnu_Omcl}, is also accurately recovered. \\

Finally, let us note that we separately tested our reconstructions in the presence of massive neutrinos without assuming the functional form of $h(z)$, given by \eqref{Hnu} but using instead the (reconstructed) \emph{effective} $\fde(z)$ in Eq. \eqref{h2}, which captures the effect of relativistic species \cite{Calderon2022}. 
Our conclusions remain unaltered, but no information on the neutrino mass can be obtained.

\section{Conclusions}
In a companion paper \citet{Calderon2022}, we jointly reconstructed the growth and expansion histories inside GR directly from the data and using minimal assumptions. We showed that our framework is able to capture a wide variety of behaviors in the DE component. In this work, we extend our methodology to include possible modifications of gravity at late times, as encoded by the function $\Geff(z)$ appearing in the (modified) Poisson equation. We illustrate the efficiency of our method in reconstructing different theories of gravity by reconstructing two phenomenological \emph{shapes} of $\mu(z)\equiv \Geff/G$. As an example, we consider a ``bump'' and a smooth transition (``dip'') towards the weak gravity regime in the recent past. We used the reconstructed $h(z)$ from background-only data, as obtained in \cite{Calderon2022} in order to fit $\fseight(z)$ to RSD mock data, thereby constraining $\mu(z)$ using minimal assumptions. We also explore the effects of incorrectly assuming a \lcdm\ background. In both cases, the fiducial $\mu(z)$ is within the $1\sigma$ confidence intervals of our reconstructions, if the background is accurately determined, and within $2\sigma$ if we incorrectly assume the \lcdm's best-fit $h(z)$. Finally, we explored the impact of massive neutrinos on the reconstructions of $\mu(z)$.
To summarize, let us list a few important results.
\begin{itemize}
    \item If the background is \emph{given} (Fig. \ref{fig:ideal_joint}), or \emph{accurately reconstructed} from SN+BAO (Fig.  \ref{fig:bkg-reco}), our reconstructions of $\Geff(z)$ are able to distinguish both fiducial $\mu$-profiles from GR at $\gtrsim2\sigma$ (see Figs. \ref{fig:ideal_joint} and \ref{fig:realistic}).
    \item Incorrectly assuming a \lcdm\ expansion (with the best-fit $\Omo$ to background probes) can lead to \emph{biased/degraded reconstructions} (red-shaded regions in Fig. \ref{fig:realistic_LCDM}) and/or \emph{biased estimations} of the amplitude of fluctuations \seighto\ (see Table \ref{tab:Constraints}). This is despite the perfect agreement with $\fseight(z)$ measurements, as shown in the lower panel of Fig. \ref{fig:realistic_LCDM}.
    \item The posterior distributions for the hyperparameters \emph{clearly} show the \emph{need} for a deviation from the mean $\bar f=1$, \emph{i.e.} GR is \emph{not} a good description of the data. This is understood because the marginalized contours in Fig. \ref{fig:posteriors_ST_vs_fR} suggest {the data are not consistent with vanishing values of $\sigma_f$, i.e. the posterior does not extend to $\sigma_f\to0$, and therefore \emph{require} deviations from the considered mean function. }Interestingly, the redshift of the transition $z_c$ is also not compatible with small values of $z_c$, and we have a ``detection'' on when this transition from $G\to\Geff$ happens; seen as a clear bump in Fig. \ref{fig:posteriors_ST_vs_fR}.
\end{itemize}  

In this work, we used forecasted (stage-IV) SN+BAO data to reconstruct the DE evolution $\fde(z)$---which determines the expansion history $h(z)$---and separately reconstructed $\mu(z)$ using DESI-like $\fseight(z)$ measurements for two different toy models of $\Geff$. We expect our methodology to hold for essentially any (viable) form of \Geff. We showed that for both profiles considered in this work, the reconstructions are able to detect the deviations from GR at $\gtrsim 2\sigma$ in the redshift range $0.5\lesssim z\lesssim1.5$ where DESI's (RSD) constraining power lies. The inclusion of external data sets, such as the (modified) luminosity distance of gravitational waves $d_L^{\rm GW}(z)$ \cite{Belgacem_2018} or the Integrated Sachs-Wolfe effect (ISW) seen in the temperature anisotropies of the cosmic microwave background (CMB) in cross-correlation with LSS surveys would provide interesting (model-independent) constraints on the allowed deviations from GR \cite{Kreisch_2018}. Moreover, we note that the effect of massive neutrinos would be 
tracked more accurately  
if we allow for a scale-dependent growth. 
We leave such extensions for future work.
 
\section*{Acknowledgements}
We thank Eric Linder for comments on the draft.
B.L. acknowledges the support of the National Research Foundation of Korea (NRF-2019R1I1A1A01063740 and NRF-2022R1F1A1076338) and the support of the Korea Institute for Advanced Study (KIAS) grant funded by the government of Korea. 
A.S. would like to acknowledge the support by National Research Foundation of Korea NRF2021M3F7A1082053, and the support of the Korea Institute for Advanced Study (KIAS) grant funded by the government of Korea.
A.A.S. was partly supported by the Project No. 0033-2019-0005 of the Russian Ministry of Science and Higher Education.

\appendix 
\section{GAUSSIAN PROCESS WITH OBSERVATIONS ON THE DERIVATIVES}\label{GP_derivative}

\begin{figure}
    \centering
    \includegraphics[width=\columnwidth]{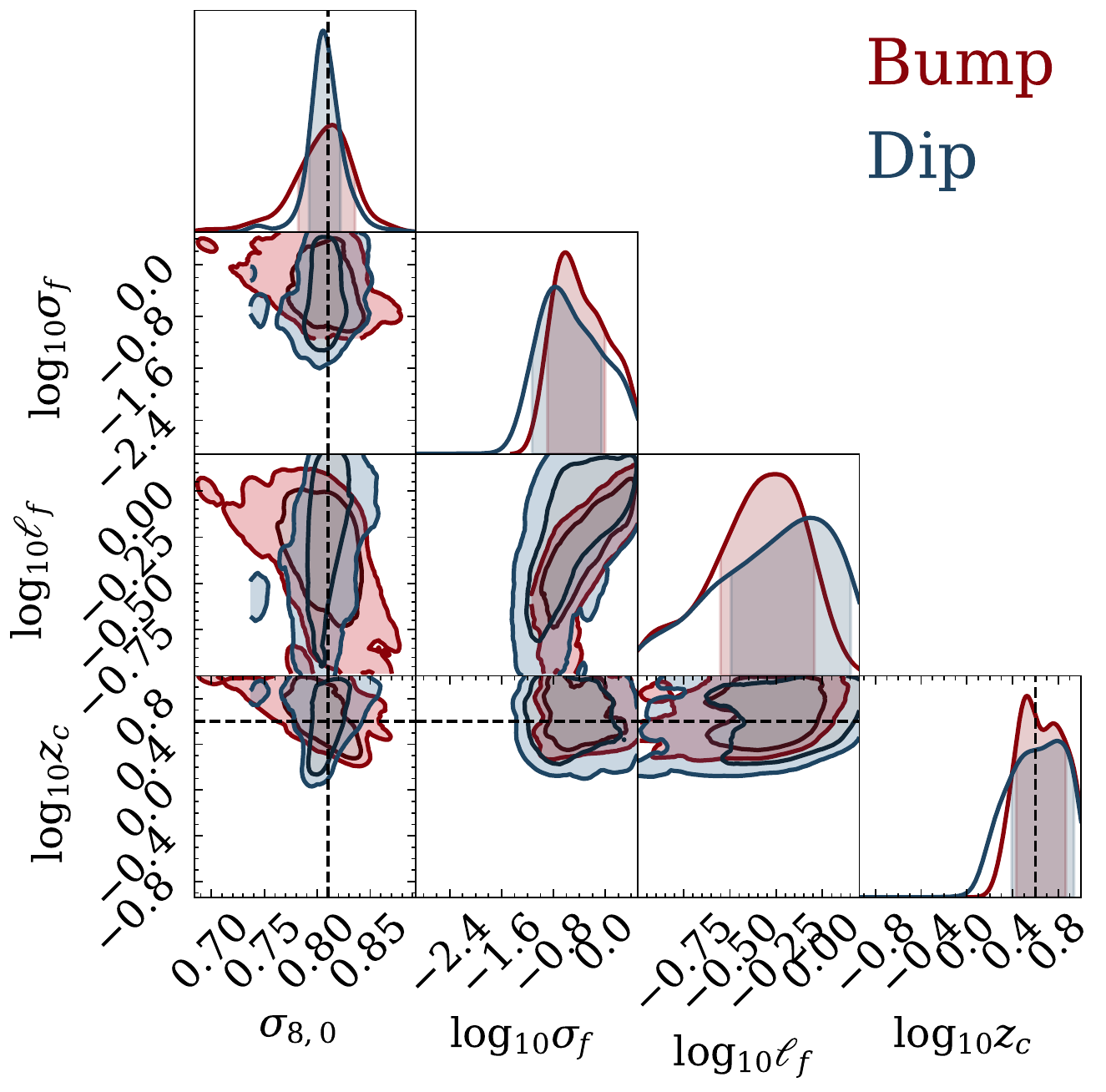}
    \caption{Marginalized posterior distributions for the relevant parameters from the RSD chains. The background expansion history used in the analysis is fixed to the median $h(z)$ obtained from the SN+BAO chains, shown as a gray line in Fig. \ref{fig:bkg-reco}.}
    \label{fig:posteriors_ST_vs_fR}
\end{figure}

\begin{table}
    \centering
    \begin{tabular}{cccc}
        \hline
		Model & $\sigma_{8,0}$ & $\log_{10}z_c$ &$\ln{\mathcal{L}}$ \\ 
		\hline
		Bump & $0.814^{+0.022}_{-0.032}$ & $0.524^{+0.342}_{-0.092}$ & $-10.15^{+0.78}_{-1.48}$ \\ 
		Dip & $0.806^{+0.017}_{-0.013}$ & $0.79^{+0.15}_{-0.40}$ &$-10.36^{+0.96}_{-1.46}$ \\ 
		Bump $(\Lambda \rm CDM)$ & $0.812^{+0.013}_{-0.021}$ & $0.86^{+0.10}_{-0.42}$ & $-10.35^{+0.94}_{-1.97}$ \\ 
		Dip $(\Lambda \rm CDM)$ & $0.784^{+0.023}_{-0.015}$ & $-0.17^{+0.20}_{-0.34}$     & $-10.5^{+1.0}_{-1.3}$ \\ 
		\hline
    \end{tabular}
    \caption{Marginalized constraints (median and $68\%$ confidence levels) on the relevant parameters. We compare the effects of incorrectly assuming \lcdm\ (bottom-two rows) vs the reconstructed $h(z)$ from SN+BAO in the upper-two rows.{In the last column, we report the log-likelihood values $\ln\mathcal{L}\equiv-\frac12\chi^2$}.}\label{tab:Constraints}
\end{table}

In this section, we describe a less common use of Gaussian process when we also observe the derivative of the function $f$ to be reconstructed \citep{rasmussen2006gaussian,2012PhRvD..85l3530S}. 
We note that in this section, $f$ denotes a general function, not the growth rate. 
In our case, $f = \mu(z)$. In addition to observations of $y$, we also ``observe'' $y' = f'(x) + \varepsilon$, where,
\begin{equation}
    \varepsilon \sim \mathcal{N}(0,C_{y'}),
\end{equation}
is a Gaussian noise and $C_{y'}$ is the covariance of the derivatives.

We further assume that $y$ and $y'$ are uncorrelated.
Therefore, the vector 
\begin{equation}
    \begin{bmatrix}
    y\\y'\\f\\f'
    \end{bmatrix}
\end{equation}
is jointly Gaussian, and the posterior predictive distribution can be calculated using
\begin{align}
    \begin{bmatrix}
    f\\
    f'
    \end{bmatrix}|y,y', X, X_* \sim
    \mathcal{N}\left(
    \begin{bmatrix}
    \bar f\\
    \bar f'
    \end{bmatrix};
    \begin{bmatrix}
    A-C B^{-1} C^T
    \end{bmatrix}
    \right),
\end{align}
where the mean is
\begin{align}
    \begin{bmatrix}
    \bar f\\
    \bar f'
    \end{bmatrix} &= C B^{-1} \begin{bmatrix}
    y-\mu_y\\
    y'-\mu_{y'}
    \end{bmatrix}, 
\end{align}
and the covariance matrix is given by
\begin{subequations}
\begin{align}   
    A & = \begin{bmatrix}
    K_{**} & K_{**}^{01} \\
    K_{**}^{10} & K_{**}^{11}\\
    \end{bmatrix} \in \mathbb{M}_{2n_*,2n_*},\\
B & = \begin{bmatrix}
K + C_y & K^{01}\\
K^{10} & K^{11} + C_{y'}
\end{bmatrix}\in \mathbb{M}_{n+n'},\\
C^T & = \begin{bmatrix}
K_* & K_*^{01} \\
K_*^{10} & K_*^{11} 
\end{bmatrix} \in \mathbb{M}_{n+n',2n_*},
\end{align}
\end{subequations}
where 
\begin{subequations}
\begin{align}   
K & = k(X,X),\\
K_* &= k(X,X_*),\\
K_{**} & = k(X_*,X_*),
\end{align}
\end{subequations}
and for any matrix $X$,
\begin{align}   
X^{i,j} &=  \frac{\partial^{i+j}X}{\partial X^i\partial X^j}. 
\end{align}
The subscript $*$ denote the set of points $X$ where the observations are done.\footnote{In our analysis, the observations are done in redshift, so that $X_*=z_*$. However, this formalism is general and can be applied to any input variable $X$.} This formalism allows us to impose theoretical priors on the samples of $\mu(z)$ and its derivative $\mu'(z)$ to smoothly recover the expected GR behavior at early times [see Eq. \eqref{Geff}].

\section{Distribution of the hyperparameters}
\label{Supplementary} 
Inspecting the posterior distributions of the hyperparameters, shown in Fig. \ref{fig:posteriors_ST_vs_fR}, can yield additional information on the $\Geff$ reconstructions and put interesting constraints on the departures from GR. First, let us note that the inferred value of $\seighto$ is unbiased in both cases, when the evolution of DE $\fde(z)$ is reconstructed using our model independent approach \cite{Calderon2022}. This is not the case when one (incorrectly) assumes a \lcdm\ expansion history (see Table \ref{tab:Constraints}).  Second, both the bump and dip reconstructions seem to require a deviation from the mean function $\bar f=\mu=1$ (\emph{i.e.} GR), as the posteriors of $\log_{10}\sigma_f$ are not compatible with $\sigma_f\to0$. This suggest that GR is not a good description of the growth $\fseight(z)$ history and that the data requires extra flexibility, as encoded by the GP kernel in Eq. \eqref{eq:kernel}. Lastly, the posteriors of $\log_{10}z_c$ seem to peak at the redshift $z_c\sim3$ where the departures from GR actually takes place (depicted by the vertical dashed line in Fig. \ref{fig:posteriors_ST_vs_fR}).  

\bibliography{_biblio}

\end{document}

%% file: defs.tex
\usepackage{graphicx}
\usepackage{dcolumn}
\usepackage{bm}
\usepackage[colorlinks=true]{hyperref}
\usepackage[mathlines]{lineno}
\usepackage{aas_macros}
\usepackage{color}
\usepackage{bbm}
\usepackage{multirow}
\usepackage{siunitx}
   
\newcommand{\chisq}{\ensuremath{\chi^2}}

\newcommand{\lcdm}{$\Lambda$CDM}

\newcommand{\Omm}{\ensuremath{\Omega_\text{m}}}
\newcommand{\Omo}{\ensuremath{\Omega_{\text{m},0}}}

\newcommand{\Omtot}{\ensuremath{\Omo^{\rm tot}}}

\newcommand{\vect}[1]{\ensuremath{\boldsymbol{#1}}}
\newcommand{\tens}[1]{\ensuremath{\mathbf{#1}}} 
\newcommand{\diff} {\ensuremath{\mathrm{d}}}

\newcommand{\abs}[1]{\ensuremath{\left\lvert#1\right\rvert}}

\newcommand{\fseight}{\ensuremath {f\sigma_8}}

\newcommand{\seighto}{\ensuremath {\sigma_{8,0}}}
\newcommand{\Geff}{\ensuremath {G_\mathrm{eff}}}
\newcommand{\fde}{\ensuremath {f_\mathrm{DE}}}

\usepackage{scalerel}
\usepackage{tikz}
\usetikzlibrary{svg.path}

\definecolor{orcidlogocol}{HTML}{A6CE39}
\tikzset{
  orcidlogo/.pic={
    \fill[orcidlogocol] svg{M256,128c0,70.7-57.3,128-128,128C57.3,256,0,198.7,0,128C0,57.3,57.3,0,128,0C198.7,0,256,57.3,256,128z};
    \fill[white] svg{M86.3,186.2H70.9V79.1h15.4v48.4V186.2z}
                 svg{M108.9,79.1h41.6c39.6,0,57,28.3,57,53.6c0,27.5-21.5,53.6-56.8,53.6h-41.8V79.1z M124.3,172.4h24.5c34.9,0,42.9-26.5,42.9-39.7c0-21.5-13.7-39.7-43.7-39.7h-23.7V172.4z}
                 svg{M88.7,56.8c0,5.5-4.5,10.1-10.1,10.1c-5.6,0-10.1-4.6-10.1-10.1c0-5.6,4.5-10.1,10.1-10.1C84.2,46.7,88.7,51.3,88.7,56.8z};
  }
}

\newcommand\orcid[1]{\href{https://orcid.org/#1}{\mbox{\scalerel*{
\begin{tikzpicture}[yscale=-1,transform shape]
\pic{orcidlogo};
\end{tikzpicture}
}{|}}}}
\def\tpdf#1{\texorpdfstring{#1}{Lg}}

\definecolor{deepmagenta}{rgb}{0.8, 0.0, 0.8}
\definecolor{ballblue}{rgb}{0.13, 0.67, 0.8}

\definecolor{RedWine}{rgb}{0.743,0,0}